\documentclass[aps,prd,twocolumn,showpacs]{revtex4}

\usepackage{graphicx,amsmath,amsfonts,amssymb,mathrsfs,epic,eepic}

\newcommand{\ket}[1]{\left| #1 \right\rangle}

\newcommand{\braket}[2]{\left\langle #1 | #2 \right\rangle}

\newcommand{\by}{\times}

\renewcommand{\d}{\mathrm{d}} % rak differential
\newcommand{\dd}[2]{\frac{\mathrm{d} #1}{\mathrm{d} #2}} % Derivata
\newcommand{\pwr}[1]{\times 10^{#1}}
\newcommand{\mtrx}[2]{\left(\begin{array}{#1}#2\end{array}\right)} % Matris

\begin{document}

\title{Day-night effect in solar neutrino oscillations with three
flavors}

\author{Mattias Blennow}\email{mbl@theophys.kth.se}
\affiliation{Division of Mathematical Physics, Department of Physics, Royal Institute of Technology (KTH), AlbaNova University Center, Roslagstullsbacken 11,
106 91 Stockholm, Sweden}

\author{Tommy Ohlsson}\email{tommy@theophys.kth.se}
\affiliation{Division of Mathematical Physics, Department of Physics, Royal Institute of Technology (KTH), AlbaNova University Center, Roslagstullsbacken 11,
106 91 Stockholm, Sweden}

\author{H{\aa}kan Snellman}\email{snell@theophys.kth.se}
\affiliation{Division of Mathematical Physics, Department of Physics, Royal Institute of Technology (KTH), AlbaNova University Center, Roslagstullsbacken 11,
106 91 Stockholm, Sweden}

\begin{abstract}
We investigate the effects of a nonzero leptonic mixing angle
$\theta_{13}$ on the solar neutrino day-night asymmetry. Using a
constant matter density profile for the Earth and well-motivated
approximations, we derive analytical expressions for the $\nu_e$
survival probabilities for solar neutrinos arriving directly at the
detector and for solar neutrinos which have passed through the
Earth. Furthermore, we numerically study the effects of a non-zero
$\theta_{13}$ on the day-night asymmetry at detectors and find that
they are small. Finally, we show that if the uncertainties in the
parameters $\theta_{12}$ and $\Delta m^2$ as well as the uncertainty
in the day-night asymmetry itself were much smaller than they are
today, this effect could, in principle, be used to determine
$\theta_{13}$.
\end{abstract}

\pacs{14.60.Pq, 13.15.+g, 26.65.+t}
%\keywords{neutrino oscillations, matter effects,
%solar neutrinos, solar neutrino day-night effect}

\maketitle

\section{Introduction}

Neutrino oscillation physics has entered the era of precision
measurements with the results from the Super-Kamiokande
\cite{Fukuda:1998mi}, SNO
\cite{Ahmad:2001an,Ahmad:2002jz,Ahmad:2002ka,Ahmed:2003kj}, and
KamLAND \cite{Eguchi:2002dm} experiments.  Especially, impressiver
results have recently come from measurements of
solar neutrinos (see
Refs.~\cite{Ahmad:2002ka,Ahmed:2003kj,Smy:2003jf}) and the solar
neutrino problem has successfully been solved in terms of solar
neutrino oscillations.

The solar neutrino day-night effect, which measures the relative
difference of the electron neutrinos coming from the Sun at nighttime
and daytime, is so far the best long baseline experiment that can
measure the matter effects on the neutrinos, the so-called
Mikheyev-Smirnov-Wolfenstein (MSW) effect
\cite{Wolfenstein:1978ue}. In all accelerator long baseline
experiments, the neutrinos cannot be made to travel through
vacuum. The atmospheric neutrino experiments, on the other hand, use
different baseline lengths for neutrinos traversing the Earth and
those that pass through vacuum. With the advent of the precision era
in neutrino oscillation physics, we can gradually hope to obtain
better measurements of the day-night effect. Recently, both
Super-Kamiokande and SNO have presented new measurements
\cite{Ahmad:2002ka,Smy:2003jf} of this effect that have errors
approaching a few standard deviations in significance.

In this paper we analyze the day-night effect in the three neutrino
flavor case. Earlier analyses of this effect, with a few exceptions,
have been performed for the two neutrino flavor case. Furthermore, the
present data also permit a new treatment of the effect due to the
particular values of leptonic mixing angles and neutrino mass squared
differences obtained from other experiments.  There are six parameters
that describe the neutrinos in the minimal extension of the standard
model: three leptonic mixing angles $\theta_{12}$, $\theta_{13}$, and
$\theta_{23}$, one \emph{CP}-phase $\delta$, and two neutrino mass
squared differences $\Delta M^2=m_3^2 - m_1^2$ and $\Delta m^2= m_2^2
- m_1^2$. The solar neutrino day-night effect is mainly sensitive to
the angles $\theta_{12}$ and $\theta_{13}$, and the mass squared
difference $\Delta m^{2}$.  Our goal is to obtain a relatively simple
analytic expression for the day-night asymmetry that reproduces the
main features of the situation. It turns out that one can come a long
way towards this goal.

Earlier treatments of the day-night effect can be found in
Refs.~\cite{Carlson:1986ui,Lisi:1997yc,Guth:1999pi,Dighe:1999id,Gonzalez-Garcia:2000dj,deHolanda:2003nj,Bandyopadhyay:2003pk}. Our
three flavor treatment is consistent with the modifications presented
by de Holanda and Smirnov \cite{deHolanda:2003nj} as well as
Bandyopadhyay {\it et al.}  \cite{Bandyopadhyay:2003pk}. 

This paper is organized as follows. In Sec.~\ref{sec:nflav} we
investigate the electron neutrino survival probability with $n$
flavors for solar neutrinos arriving at the Earth and for solar
neutrinos going through the Earth. Next, in Sec.~\ref{sec:3flav} we
study the case of three neutrino flavors, including production and
propagation in the Sun as well as propagation in the Earth. At the end
of this section, we present the analytical expression for the
day-night asymmetry. Then, in Sec.~\ref{sec:detectors} we discuss the
day-night effect at detectors. Especially, we calculate the elastic
scattering day-night asymmetry at the Super-Kamiokande experiment and
the charged-current day-night asymmetry at the SNO
experiment. Furthermore, we discuss the possibility of determining
the leptonic mixing angle $\theta_{13}$ using the day-night
asymmetry. In Sec.~\ref{sec:S&C} we present our summary as well as our
conclusions. Finally, in the Appendix we shortly review for
completeness the day-night asymmetry in the case of two neutrino
flavors.

\section{The $\boldsymbol n$ flavor solar neutrino survival probability}
\label{sec:nflav}

Assuming an incoherent neutrino flux \cite{Guth:1999pi,Dighe:1999id},
the $\nu_e$ survival probability for solar neutrinos is
\begin{equation}
\label{eq:PSn}
P_S = \sum_{i=1}^n k_i |\braket{\nu_e}{\nu_i}|^2 = \sum_{i=1}^n k_i |U_{ei}|^2,
\end{equation}
where $n$ is the number of neutrino flavors and $k_i$ is the fraction
of the mass eigenstate $\ket{\nu_i}$ in the flux of solar
neutrinos. From unitarity it follows that
\begin{equation}
\label{eq:kn}
\sum_{i = 1}^n k_i = 1.
\end{equation}
In the case of even mixing, i.e., $k_i = 1/n$ for all $i$, we obtain
$P_S = 1/n$.

For neutrinos reaching the Earth during daytime (at the detector
site), $P_S$ is the $\nu_e$ survival probability at the
detector. However, during nighttime this survival probability may be
altered by the influence of the effective Earth matter density
potential.  Thus, in this case, the survival probability becomes
\begin{equation}
\label{eq:PSEn}
P_{SE} = \sum_{i=1}^n k_i |\braket{\nu_e}{\tilde\nu_i}|^2,
\end{equation}
where $\ket{\tilde{\nu_i}} = \ket{\nu_i(L)}$ and $L$ is the length of the
neutrino path through the Earth. Here, the components of
$\ket{\nu_i(t)}$ satisfy the Schr\"odinger equation
\begin{equation}
\label{eq:diffeq}
\mathrm{i}\dd{\ket{\nu_i(t)}_m}{t} = H_m \ket{\nu_i(t)}_m
\end{equation}
with the initial condition $\ket{\nu_i(0)} = \ket{\nu_i}$ and where
$m$ denotes the mass eigenstate basis.

The Hamiltonian $H$ is given by (assuming $k$ sterile neutrino
flavors)
\begin{equation}
H_m \simeq \frac{M^2}{2p} + 
U^\dag {\rm diag}(V_{CC}, 0, \ldots, 0, 
\overbrace{-V_{NC}, \ldots, -V_{NC}}^{k \ {\rm times}})U,
\end{equation}
where $M = {\rm diag}(m_1,m_2,\ldots,m_n)$, the effective
charged-current Earth matter density potential is $V_{CC} =
\sqrt{2}G_F N_e$, and the effective neutral-current Eath matter
density potential is $V_{NC} = -G_F N_n / \sqrt{2}$, where $G_F$ is
the Fermi coupling constant and where $N_e$ and $N_n$ are the electron
and nucleon number densities, respectively. The number densities are
functions of $t$ depending on the Earth matter density profile, which
is normally given by the Preliminary Reference Earth Model (PREM)
\cite{Dziewonski:1981xy}. The term
$|\braket{\nu_e}{\tilde{\nu_i}}|^2$ is interpreted as the probability
of a neutrino reaching the Earth in the mass eigenstate $\ket{\nu_i}$
to be detected as an electron neutrino after traversing the distance
$L$ in the Earth. For notational convenience we denote
\begin{equation}
P_{ie} = |\braket{\nu_e}{\tilde{\nu_i}}|^2.
\end{equation}
Clearly, $P_{ie}(L=0) = |\braket{\nu_e}{\nu_i}|^2 =
|U_{ei}|^2$. Furthermore, from unitarity it follows that
\begin{equation}
\label{eq:Pien}
\sum_{i = 1}^n P_{ie} = 1.
\end{equation}
Again, in the case of even mixing, we obtain $P_{SE} = 1/n$, and the
$\nu_e$ survival probability is unaffected by the passage
through the Earth.

\section{The case of three neutrino flavors}
\label{sec:3flav}

Until now most analyses of the day-night effect have been done in the
framework of two neutrino flavors. However, we know that there are (at
least) three neutrino flavors. The reason for using two flavor
analyses has been that the leptonic mixing angle $\theta_{13}$ is
known to be small \cite{Apollonio:1999ae}, leading to an approximate
two neutrino case. One of the main goals of this paper is to find the
effects on the day-night asymmetry induced by using a non-zero mixing
angle $\theta_{13}$. In what follows, we assume that there are three
active neutrino flavors and no sterile neutrinos.

We will use the standard parametrization of the $3\by 3$ leptonic
mixing matrix \cite{Hagiwara:2002fs}
\begin{equation}
U = \mtrx{ccc}{c_{13}c_{12} & c_{13}s_{12} & s_{13}{\rm e}^{-{\rm i}\delta} \\
* & * & * \\
* & * & *},
\end{equation}
where $s_{ij} = \sin \theta_{ij}$ and $c_{ij} = \cos \theta_{ij}$,
$\theta_{ij}$ are leptonic mixing angles, and the elements denoted by
$*$ do not affect the neutrino oscillation probabilities, which we are
calculating in this paper.

\subsection{Production and propagation in the Sun}

In the three flavor framework, there are a number of issues of the
neutrino production and propagation in the Sun, which are not present
in the two flavor framework. First of all, the three energy levels of
neutrino matter eigenstates in general allow two MSW
resonances. Furthermore, the matter dependence of the mixing parameters
are far from as simple as in the two flavor case. The result of this
is that we have to make certain approximations.

Repeating the approach made in the two flavor case (see
the Appendix), we obtain the following expression for $k_i$:
\begin{equation}
k_i = \int_0^{R_\odot} \d r f(r) \sum_{j = 1}^3 |\hat U_{ej}|^2 P_{ji}^s,
\end{equation}
where $\hat U$ is the mixing matrix in matter, $P_{ji}^s$ is the
probability of a neutrino created in the matter eigenstate
$\ket{\nu_{M,j}}$ to exit the Sun in the mass eigenstate
$\ket{\nu_i}$, and $f(r)$ is the normalized spatial production
distribution in the Sun.

The second resonance in the three flavor case occurs at $V_{CC} \simeq
\cos(2\theta_{13}){\Delta M^2}/{(2E)}$, assuming that the resonances
are fairly separated. The maximal electron number density in the Sun,
according to the standard solar model (SSM) \cite{Bahcall:2000nu}, is
about $N_{e,\rm max} \simeq 102$ $N_A$/cm$^3$, yielding a maximal
effective potential $V_{CC,\mathrm{max}} \simeq 7.8\pwr{-18}$
MeV. Assuming the large mass squared difference $\Delta M^2$ to be of
the order of the atmospheric mass squared difference ($|\Delta
m_{\mathrm{atm}}^2| \simeq 2\pwr{-3}$ eV$^2$ \cite{Maltoni:2003da}),
the neutrino energy $E$ to be of the order of $10$ MeV, and
$\theta_{13}$ to be small, we find $\left|{\Delta
M^2}/{2E}\right|\cos(2\theta_{13}) \simeq 10^{-16}$ MeV. Thus, the
solar neutrinos never pass through the second resonance, independent
of the sign of the large mass squared difference.

Since the neutrinos never pass through the second resonance, it is a good
approximation to assume that the matter eigenstate $\ket{\nu_{M,3}}$
evolves adiabatically, and thus, we have
\begin{equation}
P_{3k}^s = P_{k3}^s = \delta_{3k}.
\end{equation}
Unitarity then implies that
\begin{equation}
P_{12}^s = P_{21}^s = P_{\rm jump}.
\end{equation}
Furthermore, if we assume that $V_{CC} \lesssim {\Delta m^2}/{(2E)}
\ll {\Delta M^2}/{(2E)}$, the neutrino evolution is well approximated
by the energy eigenstate $\ket{\nu_{M,3}}$ evolving as the mass
eigenstate $\ket{\nu_3}$ and the remaining neutrino states oscillating
according to the two flavor case with the effective potential
$V_{\mathrm{eff}} = c_{13}^2 V_{CC}$. This does not change the
probability $P_{\rm jump}$ as calculated with a linear approximation
of the potential in the two flavor case. This means that we may use
the same expression as that obtained in the two flavor case, see the
Appendix, even if the resonance point, where $|{N_e}/{\dot
N_e}|$ is to be evaluated, does change. However, in the Sun, $N_e$ is
approximately exponentially decaying with the radius of the Sun,
leading to $|{N_e}/{\dot N_e}|$ being approximately constant, and
thus, independent of the point of evaluation. For the large mixing
angle (LMA) region, the probability $P_{\rm jump}$ of a transition
from $\ket{\nu_{M,1}}$ to $\ket{\nu_{M,2}}$, or vice versa, is
negligibly small ($P_{\rm jump} < 10^{-1700}$). However, we keep it in
our formulas for completeness.

To make one further approximation, as long as the Sun's effective
potential is much less than the large mass squared difference $\Delta
M^2$, the mixing angle $\hat\theta_{13} \simeq \theta_{13}$ giving
\begin{equation}
k_3 \simeq \int_0^{R_\odot} \d r f(r) \sin^2\theta_{13} = \sin^2\theta_{13}.
\end{equation}

A general parametrization for $k_1$ and $k_2$ is then given by
\begin{equation}
k_1 = c_{13}^2 \frac{1+D_{3\nu}}2, \ \ k_2 = c_{13}^2 \frac{1-D_{3\nu}}2.
\end{equation}
In the above approximation, the oscillations between the matter
eigenstates $\ket{\nu_{M,1}}$ and $\ket{\nu_{M,2}}$ are well
approximated by a two flavor oscillation, using the small mass
difference squared $\Delta m^2$, the mixing angle $\theta_{12}$, and
the effective potential $V_{\mathrm{eff}} = c_{13}^2 V_{CC}$. Thus, we
obtain
\begin{equation}
\label{eq:D3nu}
D_{3\nu} = \int_0^{R_\odot} 
\d r f(r) \cos[2\hat\theta_{12}(r)] (1-2P_{\rm jump} ),
\end{equation}
where $\cos[2\hat\theta_{12}(r)]$ is calculated in the same way as in
the two flavor case using the effective potential.  For reasonable
values of the neutrino oscillation parameters, this turns out to be an
excellent approximation.

Inserting the above approximation into Eq.~(\ref{eq:PSn}) with $n =
3$, we obtain
\begin{equation}
P_S = s_{13}^4 + c_{13}^4\frac{1+D_{3\nu}\cos(2\theta_{12})}2.
\end{equation}
When $\theta_{13} \rightarrow 0$, we have $D_{3\nu} \rightarrow
D_{2\nu}$, and thus, we recover the two flavor survival probability in
this limit.

\subsection{Propagation in the Earth} 

As in the case of propagation in the Sun, $V_{CC} \lesssim {\Delta
m^2}/{2E} \ll {\Delta M^2}/{2E}$, $\ket{\nu_{M,3}} \simeq
\ket{\nu_3}$, and the remaining two neutrino eigenstates evolve
according to the two flavor case with an effective potential of
$V_{\mathrm{eff}} = c_{13}^2 V_{CC}$. For the MSW solutions of the
solar neutrino problem along with the assumption that $\Delta M^2$ is
of the same order of magnitude as the atmospheric mass squared
difference, this condition is well fulfilled for solar neutrinos
propagating through the Earth.  As a direct result, we obtain the
probability $P_{3e}$ as
\begin{equation}
P_{3e} \simeq |\braket{\nu_e}{\nu_3}|^2
= |U_{e3}|^2 = s_{13}^2.
\end{equation}
It also follows that
\begin{equation}
P_{2e} = c_{13}^4\frac{K V_E}{4a^2}\sin^2(2\theta_{12})\sin^2(aL) 
+c_{13}^2 s_{12}^2,
\end{equation}
where
\begin{equation}
a = \frac 12 \sqrt{K^2 - 2c_{13}^2 V_E K \cos(2\theta_{12}) + c_{13}^4 V_E^2},
\end{equation}
$V_E$ is the electron neutrino potential in the Earth, and
$K = {\Delta m^2}/{2E}$. We observe that when $L = 0$ or $V_E
= 0$, $P_{2e} = c_{13}^2 s_{12}^2 = |U_{e2}|^2$ just as expected.

\subsection{The final expression for $\boldsymbol{P}_{\boldsymbol{n-d}}$}

Now, we insert the analytical expressions obtained in the previous two
sections into Eq.~(\ref{eq:PSEn}) with $n=3$ and subtract $P_S$ from this in
order to obtain an expression for $P_{n-d} = P_{SE}-P_S$ in the three
flavor framework. After some simplifications, we find
\begin{equation}
\label{eq:reg3expr}
P_{n-d} = -c_{13}^6 D_{3\nu}\frac{K V_E}{4a^2} \sin^2(2\theta_{12}) \sin^2(aL).
\end{equation}
For the MSW solutions of the solar neutrino problem, $K \gg
c_{13}^2V_E$, and thus, $a \simeq  K/2$. This yields
\begin{equation}
\label{eq:reg3appr}
P_{n-d} \simeq -2 c_{13}^6 D_{3\nu}\frac{E V_E}{\Delta m^2} 
\sin^2(2\theta_{12}) \sin^2\left({\textstyle \frac{\Delta m^2}{4E}L}\right).
\end{equation}
Apparently, the effect of using three flavors instead of two is, up to
the approximations made, a multiplication by $c_{13}^6$ as well as a
correction in changing $D_{2\nu}$ to $D_{3\nu}$. When $\theta_{13}
\rightarrow 0$, we have $D_{3\nu} \rightarrow D_{2\nu}$, and we regain
the two flavor expression in this limit (see the Appendix).  An
important observation is that the regenerative term, for $V_E \ll K$,
is linearly dependent on $V_E$. Thus, the choice of which value of
$V_E$ to use is crucial for the quantitative result. As is argued in
the appendix, the potential to use is the potential corresponding to
the electron number density of the Earth's crust. However, the
qualitative behavior of the effect of a non-zero $\theta_{13}$ is not
greatly affected.

\section{The day-night effect at detectors}
\label{sec:detectors}

From the calculations made in the previous parts of this paper, we
obtain the day-night asymmetry of the electron neutrino flux at
the neutrino energy $E$ as
\begin{eqnarray}
A_{n-d}^{\phi_e}(E) &=& 2\frac{\phi_{e,N}(E)-\phi_{e,D}(E)}
{\phi_{e,N}(E)+\phi_{e,D}(E)} \nonumber \\
&=& 
\frac{P_{n-d}(E)}{P_S(E)+\frac{P_{n-d}(E)}{2}}.
\end{eqnarray}
However, this is \emph{not} the event rate asymmetry measured at
detectors. We will assume a water-Cherenkov detector in which
neutrinos are detected by one of the following reactions~\footnote{For
the neutral-current (NC) reaction, $\nu_x + d \rightarrow p + n +
\nu_x$, the cross sections for all neutrino flavors are the same to
leading order in the weak coupling constant. As a result, there will
be no day-night effect in the NC reaction.}:
\begin{eqnarray}
\nu_x + e^- & \longrightarrow & \nu_x + e^-, \\
\nu_x + d & \longrightarrow & p+p+e^-, \label{eq:CCreaction}
\end{eqnarray}
where $x = e, \mu, \tau$, which are referred to as elastic scattering
(ES) and charged-current (CC), respectively. The CC reaction can only
occur for $x=e$, since inserting $x \neq e$ in
Eq.~(\ref{eq:CCreaction}) would violate the lepton numbers $L_e$ and
$L_x$. We assume that the scattered electron energy $T'$ is measured
and that the cross sections ${\d \sigma_{\nu_\mu}}/{\d T'}$ and ${\d
\sigma_{\nu_\tau}}/{\d T'}$ are equal.

If we denote the zenith angle, i.e., the angle between zenith and the
Sun at the detector, by $\alpha$, then the event rate of measured
electrons with energy $T$ in the detector is proportional to
\begin{eqnarray}
R(\alpha,T) &=&
\int_0^\infty \d E \phi(E) \int_0^{T'_{\mathrm{max}}} \d T' \nonumber \\
&&\times F(T,T') \dd{\sigma_{\nu_{\mathrm{solar}}}}{T'},
\end{eqnarray}
where $\phi(E)$ is the total solar neutrino flux, $T'$ is the true
electron energy, and ${\d \sigma_{\nu_{\mathrm{solar}}}}/{\d T'}$ is given
by
\begin{equation}
\dd{\sigma_{\nu_{\mathrm{solar}}}}{T'} =
P_{SE}\dd{\sigma_{\nu_e}}{T'} + (1-P_{SE})\dd{\sigma_{\nu_\mu}}{T'}.
\end{equation}
Here, we have used the assumption ${\d \sigma_{\nu_\mu}}/{\d T'} =
{\d \sigma_{\nu_\tau}}/{\d T'}$, since neutrinos not found in the state
$\ket{\nu_e}$ are assumed to be in the state $\ket{\nu_\mu}$ or in the
state $\ket{\nu_\tau}$. The energy resolution of the detector is
introduced through $F(T, T')$, which is given by
\begin{equation}
F(T,T') =
\frac{1}{\Delta_{T'}\sqrt{2\pi}}
\exp\left(-\frac{(T-T')^2}{2\Delta_{T'}^2}\right),
\end{equation}
where $\Delta_{T'}$ is the energy resolution at the electron energy $T'$.

\begin{figure}
\begin{center}
\vspace{0.4cm}
\includegraphics[scale=0.3]{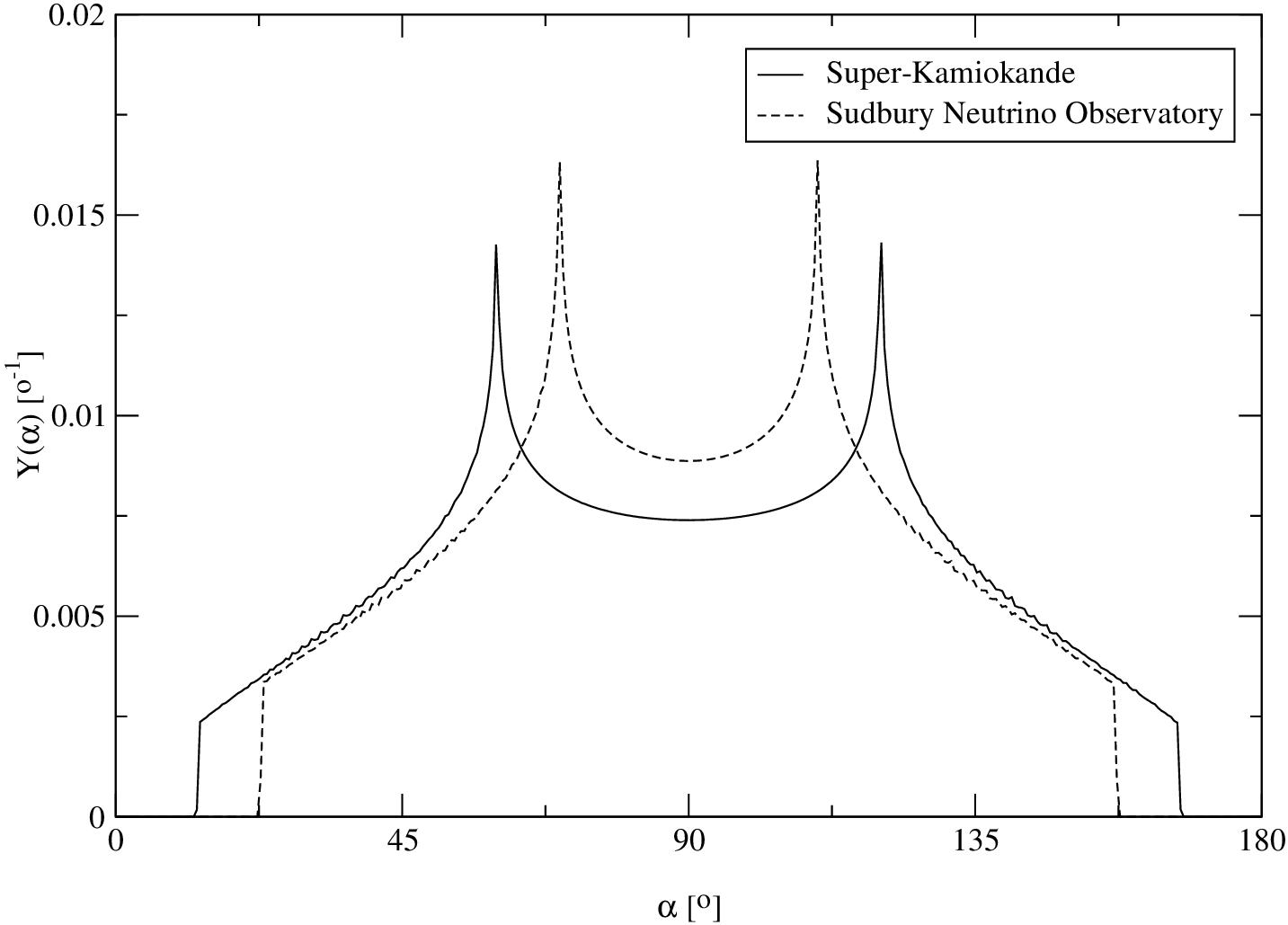}
\caption{The zenith angle exposure function $Y(\alpha)$ for SK
and SNO as a function of the zenith angle $\alpha$. The data are
retrieved from Ref.~\protect\cite{Bahcall}.}
\label{fig:exposure}
\end{center} \end{figure}

The night and day rates $N$ and $D$ at the measured electron energy
$T$ are given by
\begin{eqnarray}
D(T) &=& \int_0^{ \pi / 2} \d\alpha R(\alpha,T) Y(\alpha), \\
N(T) &=& \int_{\pi /2}^{\pi} \d\alpha R(\alpha,T) Y(\alpha),
\end{eqnarray}
respectively. Here, $Y(\alpha)$ is the zenith angle exposure function,
which gives the distribution of exposure time for the different
zenith angles. The exposure function is clearly symmetric around
$\alpha = \pi /2$ and is plotted in Fig.~\ref{fig:exposure} for
both Super-Kamiokande (SK) and SNO.
From the night and day rates at a specific electron energy, we define
the day-night asymmetry at energy $T$ as
\begin{equation}
A_{n-d}(T) = 2\frac{N(T)-D(T)}{N(T)+D(T)}.
\end{equation}
The final day-night asymmetry is given by integrating the day and
night rates over all energies above the detector threshold
energy $T_{\rm th}$, i.e.,
\begin{equation}
A_{n-d} = 2\frac{\int_{T_{\mathrm{th}}}^\infty \d T [N(T)-D(T)]}
{\int_{T_{\mathrm{th}}}^\infty \d T [N(T)+D(T)]} = 
 2\frac{N-D}{N+D}.
\end{equation}
The threshold energy $T_{\mathrm{th}}$ is 5 MeV for both SK and SNO.

For computational reasons, we will start by performing the integral
over the zenith angle $\alpha$. For the daytime flux $D$, $P_{SE} =
P_S$, which is independent of $\alpha$. As a result, the only $\alpha$
dependence is in $Y(\alpha)$ and the zenith angle integral only
contributes with a factor one-half [if the normalization of $Y$ is
such that $\int_0^\pi \d \alpha Y(\alpha) = 1$]. In order to be able to
use the results we have obtained for $P_{n-d}$, we need to compute the
difference between the night and day fluxes, which is given by
\begin{equation}
N(T)-D(T) = \int_{\pi /2}^\pi \d\alpha Y(\alpha)R^{n-d},
\end{equation}
where the quantity $R^{n-d} = R(\alpha,T)-R(\pi-\alpha,T)$ is on
the form
\begin{eqnarray}
R^{n-d} &=& \int_0^\infty \d E_\nu \phi(E_\nu) \int_0^{T'_{\mathrm{max}}}
\d T' \nonumber \\
&&\times  F(T,T') \dd{\sigma_{\nu_{\mathrm{sol}}}^{n-d}}{T'}
\end{eqnarray}
and
\begin{equation}
\dd{\sigma_{\nu_{\mathrm{sol}}}^{n-d}}{T'} =
P_{n-d}(\alpha, E_\nu) 
\left(\dd{\sigma_{\nu_e}}{T'}-\dd{\sigma_{\nu_\mu}}{T'}\right).
\end{equation}
Note that the $\alpha$ dependence in $P_{n-d}$ enters through the
length traveled by the neutrinos in the Earth and that the argument
$aL$ of the second $\sin^2$ factor in Eq.~(\ref{eq:reg3expr})
oscillates very fast and performs an effective averaging of $P_{n-d}$
in the zenith angle integral, i.e., replacing $\sin^2(aL)$ by
$1/2$. After this averaging, the only zenith angle dependence left is
that of $Y(\alpha)$ and the zenith angle integral only gives us a
factor of one-half as in the case of the day rate $D$.

\subsection{Elastic scattering detection}

Neutrinos are detected through ES at both SK and SNO. The ES
cross sections in the laboratory frame are given by
Ref.~\cite{Hagiwara:2002fs}. For kinematical reasons, the maximal
kinetic energy of the scattered electron in the laboratory frame is
given by
\begin{equation}
T'_{\mathrm{max}} = \frac{E_\nu}{1+\frac{m_e}{2 E_\nu}}.
\end{equation}

The integrals that remain cannot be calculated analytically. Hence, we
use numerical methods to evaluate these integrals. However, computing
all integrals by numerical methods demands a lot of computer time, and
thus, we make one further approximation, that all solar $^8$B
neutrinos are produced where the solar effective potential is $V_{CC}
\simeq 7.07\pwr{-18}$ MeV, which is the effective potential at the radius
where most solar neutrinos are produced. For reasonable values of the
fundamental neutrino parameters, the error made in this approximation
is small.

For the energy resolution of SK, we use \cite{Guth:1999pi}
\begin{equation}
\Delta_{T'} = 1.6 \ \mathrm{MeV} \sqrt{T'/(10 \ \mathrm{MeV})},
\end{equation}
and for the electron number density in the Earth, we use $N_e = 1.4
N_A$/cm$^3$, where $N_A$ is the Avogadro constant, which roughly
corresponds to 2.8 g/cm$^3$ (using $Z/A \simeq 0.5$, where $Z$ is the
number of protons and $A$ the number of nucleons for the mantle of the
Earth). The electron number density used corresponds to the density in
the Earth's crust.  The motivation for using this density rather than
a mean density can be found in the Appendix. Note that the
regenerative term $P_{n-d}$ in Eq.~(\ref{eq:reg3appr}), and thus, the
day-night asymmetry, is linearly dependent on the matter potential
$V_E$. It follows that the electron number density used has a great
impact on the final results. If we had used the average mantle matter
density of about 5 g$/$cm$^3$, then the resulting asymmetry would
increase by almost a factor of two.

\begin{figure}
\begin{center}
\vspace{0.4cm}
\includegraphics[scale=0.3]{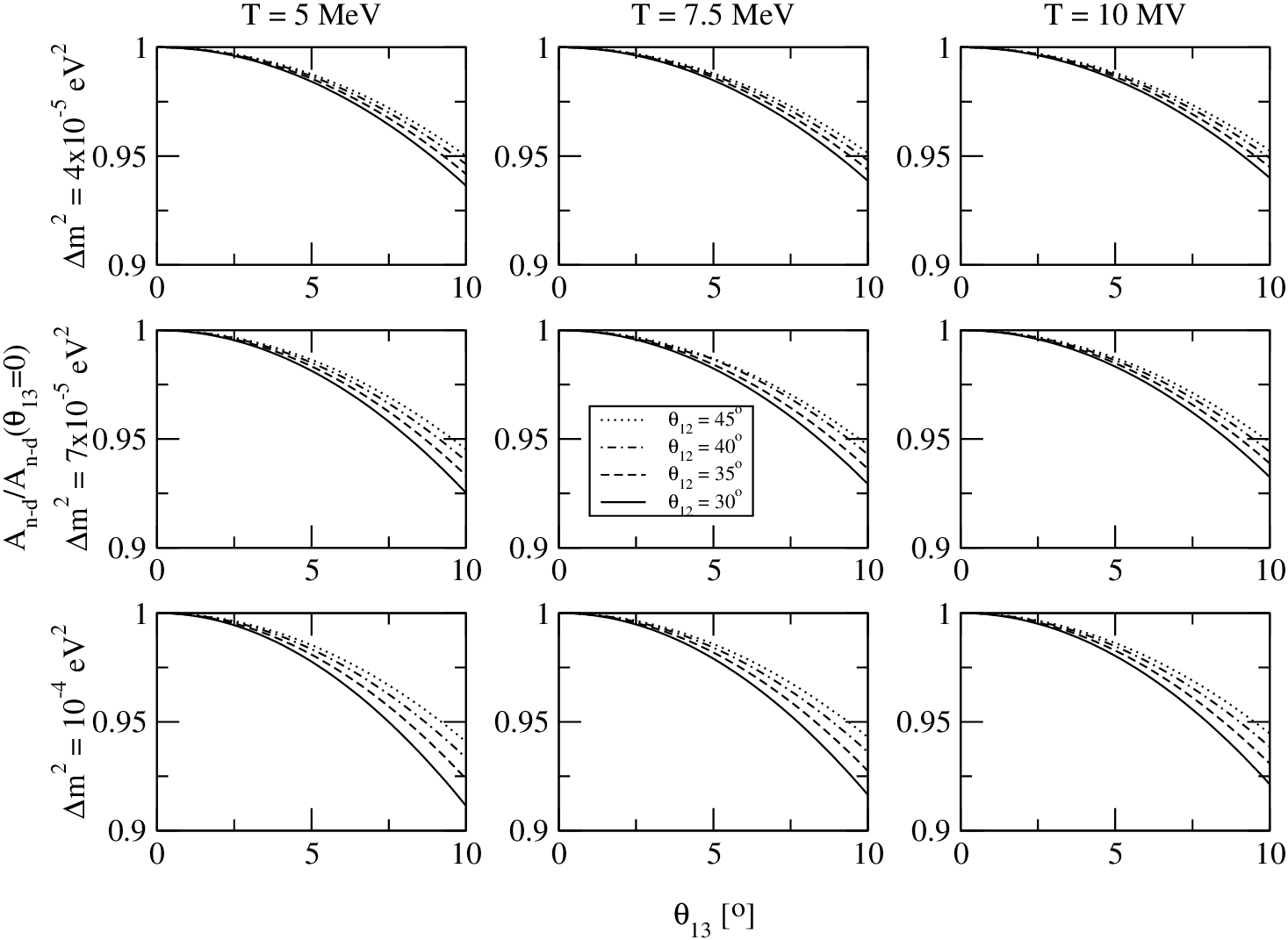}
\caption{The day-night asymmetry at SK for different values of $T$, 
$\Delta m^2$, and $\theta_{12}$ as a function of $\theta_{13}$ relative
to the corresponding value for $\theta_{13}=0$.}
\label{fig:SKres}
\end{center}
\end{figure}

\begin{figure}[!t]
\begin{center}
\vspace{0.4cm}
\includegraphics[scale=0.3]{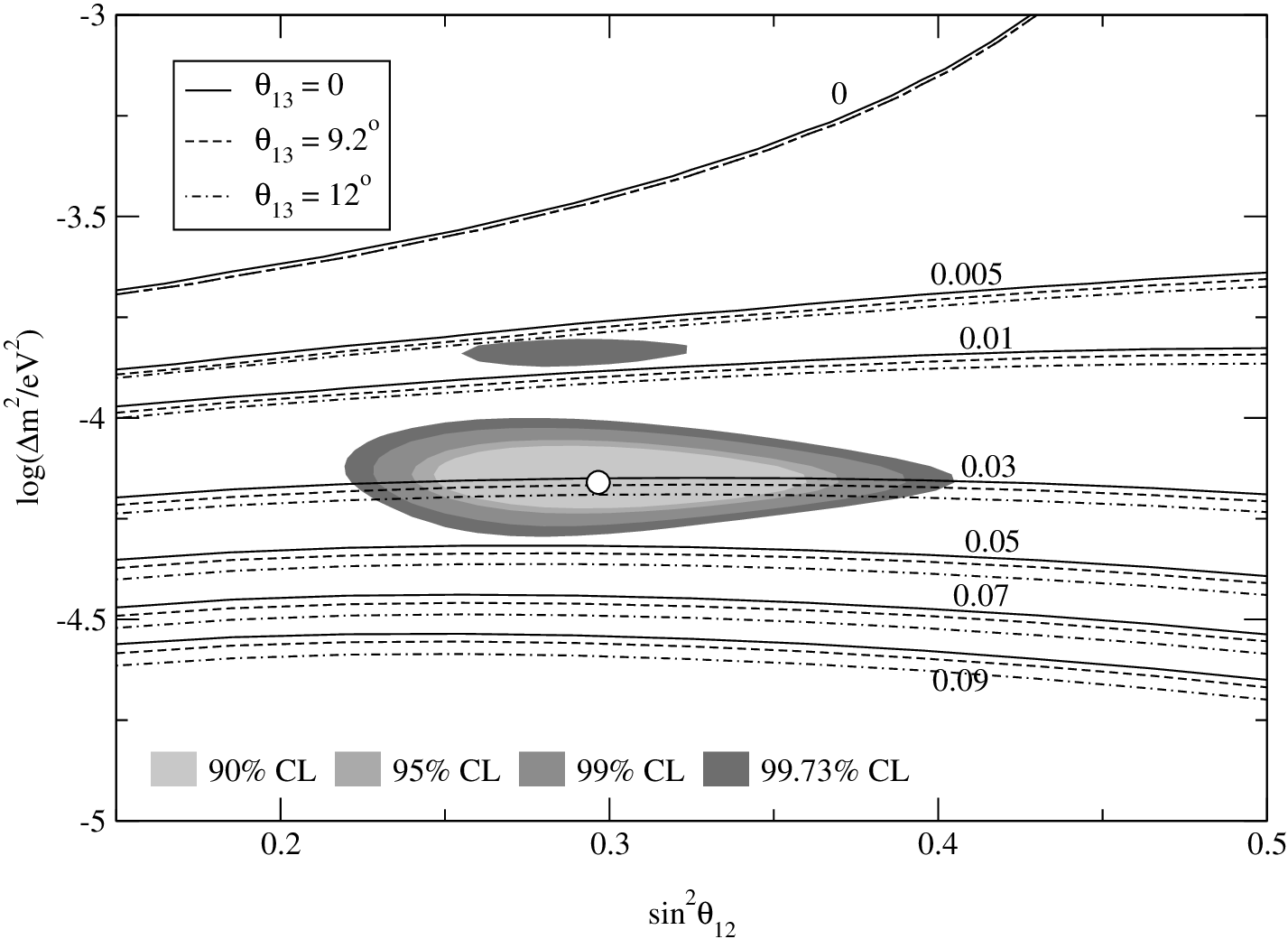}
\caption{Isocontours in the $\theta_{12}$-$\Delta m^2$ 
parameter space for the ES day-night asymmetry for three different 
values of $\theta_{13}$. The values of $A_{n-d}$ for the different 
isocontours are shown in the
figure. The shaded regions correspond to the allowed regions 
of the parameter space for different confidence
levels and the circle corresponds to the best-fit point according to
Ref.~\protect\cite{Maltoni:2003da}.}
\label{fig:EScontour}
\end{center}
\end{figure}

The above values give us the numerical results presented in
Fig.~\ref{fig:SKres}.
As can be seen from this figure, the relative effect of a non-zero
$\theta_{13}$ is increasing if the small mass squared difference
$\Delta m^2$ increases or if the measured electron energy or the
leptonic mixing angle $\theta_{12}$ decreases. The effect of changing
$\theta_{12}$ is also clearly larger for smaller electron energy $T$
and larger small mass squared difference $\Delta m^2$.  In
Fig.~\ref{fig:EScontour} the isocontours of constant day-night
asymmetry in the SK detector with $\theta_{13}$ equal to 0,
$9.2^\circ$ and $12^\circ$ are shown for a parameter space covering
the LMA solution of the solar neutrino problem. The values used for
$\theta_{13}$ correspond to no mixing as well as the CHOOZ upper bound
for $\Delta M^2$ equal to $2.5\pwr{-3}$ eV$^2$ and $2.0\pwr{-3}$
eV$^2$, respectively \cite{Apollonio:2002gd}.
As can be seen from this figure, the variation in the isocontours are
small compared to the size of the LMA solution and to the current
uncertainty in the day-night asymmetry
\cite{Ahmad:2002ka,Smy:2003jf}. However, if the values of the
parameters $\theta_{12}$, $\Delta m^2$, and $A_{n-d}$ were known with
a larger accuracy, then the change due to non-zero $\theta_{13}$
could, in principle, be used to determine the ``reactor'' mixing angle
$\theta_{13}$ as an alternative to long baseline experiments such as
neutrino factories \cite{Huber:2002mx,Apollonio:2002en} and
super-beams \cite{Huber:2002mx,Huber:2003pm,Itow:2001ee} as well as
future reactor experiments \cite{Huber:2003pm} and matter effects for
supernova neutrinos \cite{Dighe:2003jg}. The day-night asymmetry for
the best-fit value of Ref.~\cite{Maltoni:2003da} is $A_{n-d} \simeq
3.0$ \%, which is larger than the theoretical value quoted by the SK
experiment, but still clearly within one standard
deviation of the experimental best-fit value $A_{n-d} = (1.8\pm 1.6
{}^{+1.2}_{-1.3})$ \% \cite{Smy:2003jf}.

\subsection{Charged-current detection}

Only SNO uses heavy water, and thus, SNO is the only experimental
facility detecting solar neutrinos through the CC reaction
(\ref{eq:CCreaction}). Since the electron mass $m_e$ is much smaller
than the proton mass $m_p$ ($m_e \simeq 511$ keV, $m_p \simeq 938$
MeV), most of the kinetic energy in the center-of-mass frame, which is
well approximated by the laboratory frame, since the deuteron mass by far
exceeds the neutrino momentum, after the CC reaction will be carried
away by the electron. This energy is given by
\begin{equation}
T' = E + \Delta E_{\mathrm{mass}},
\end{equation}
where $\Delta E_{\mathrm{mass}} = m_d - 2m_p - m_e \simeq - 1.95$ MeV. Thus,
we approximate the differential cross section
${\d \sigma_{\nu_e}}/{\d T'}$ by
\begin{equation}
\label{eq:crossCCe}
\dd{\sigma_{\nu_e}}{T'} = \sigma_{\nu_e} \delta(T'-E+1.95 \ \mathrm{MeV}).
\end{equation}
In the above expression, we use the numerical results given in
Ref.~\cite{Ando:2002pv} and perform linear interpolation to calculate the
total cross section $\sigma_{\nu_e}$ as a function of the neutrino
energy $E$. For $x \neq e$, the reaction in
Eq.~(\ref{eq:CCreaction}) is forbidden, since it violates the lepton
numbers $L_e$ and $L_x$. Thus, for $x \neq e$, we have
${\d \sigma_{\nu_x}}/{\d T'} = 0$.

\begin{figure}
\begin{center}
\vspace{0.35cm}
\includegraphics[scale=0.3]{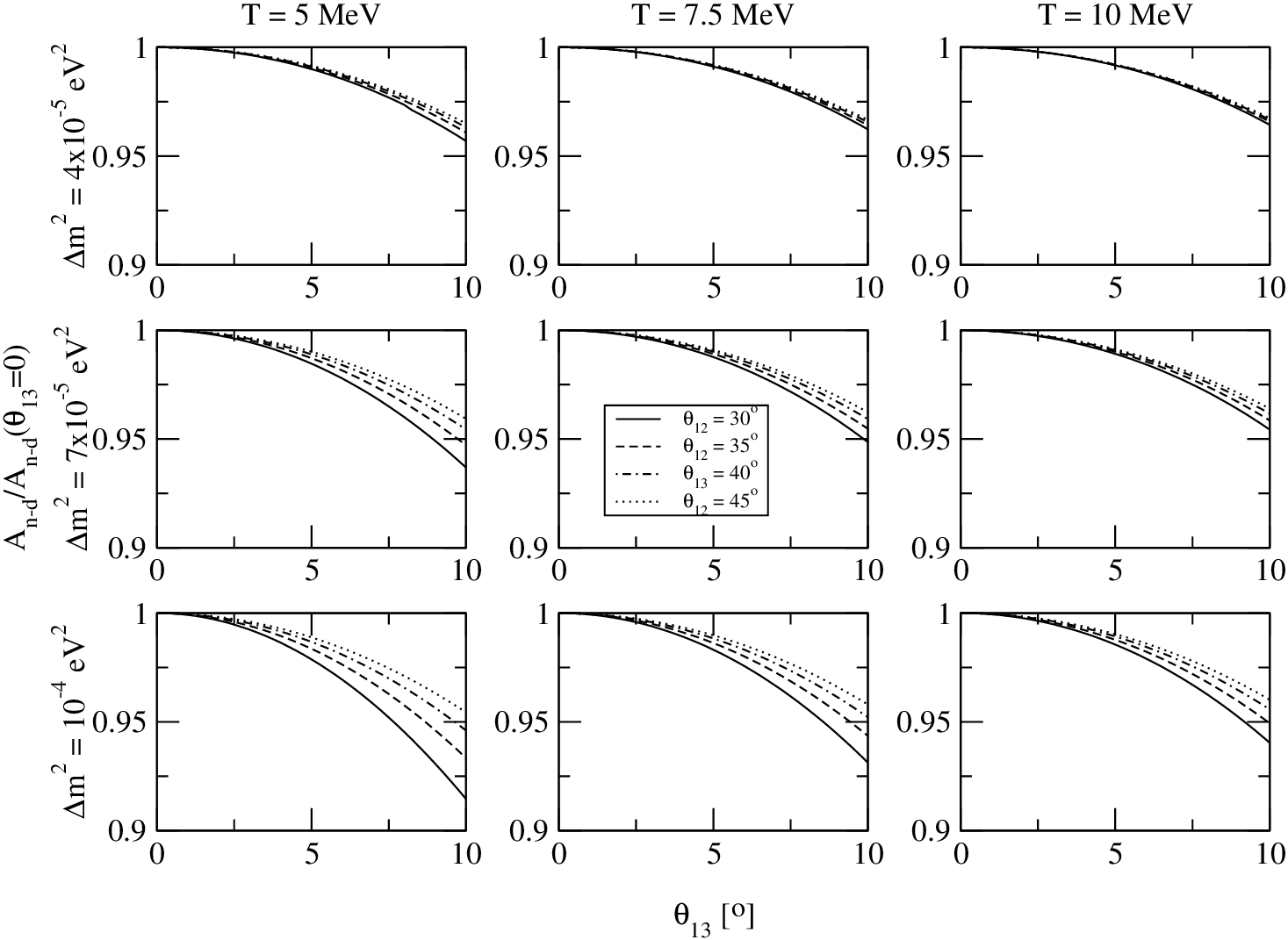}
\caption{The CC day-night asymmetry at SNO for different values of $T$, 
$\Delta m^2$, and $\theta_{12}$ as a function of $\theta_{13}$ relative
to the  corresponding value for $\theta_{13}=0$.}
\label{fig:SNOres}
\end{center}
\end{figure}

The energy resolution at SNO is given by \cite{Ahmad:2002jz,SNOHOWTO}
\begin{eqnarray}
\Delta_{T'} &=& -0.0684 \ \mathrm{MeV} \nonumber \\
&&+ 0.331 \ \mathrm{MeV} \ {\textstyle \sqrt{\left({T'}/{\rm MeV}\right)}}
\nonumber \\
&&+ 0.0425 \ {\rm MeV} \ {\textstyle \left({T'}/{\rm MeV}\right)}.
\end{eqnarray}
This gives the results presented in Fig.~\ref{fig:SNOres}.
This figure shows the same main features as Fig.~\ref{fig:SKres}.
However, the effects of different $T$ and $\Delta m^2$ are larger in
the CC case.

\begin{figure}
\begin{center}
\vspace{0.2cm}
\includegraphics[scale=0.3]{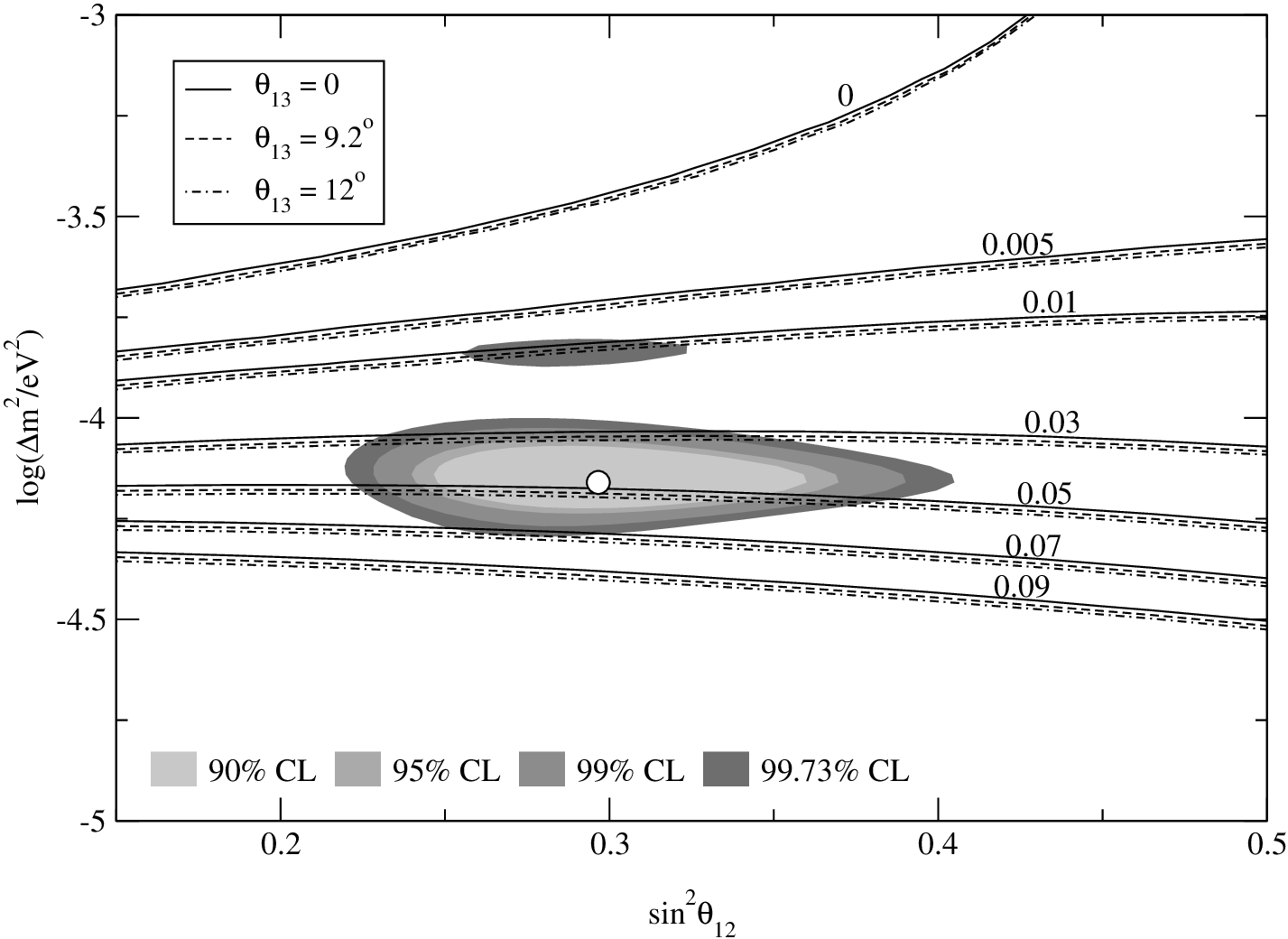}
\caption{Isocontours for the CC day-night asymmetry in the
$\theta_{12}$-$\Delta m^2$ parameter space for three different values
of $\theta_{13}$.  The values of $A_{n-d}$ for the different
isocontours are shown in the figure. The shaded regions and the circle
are the same as in Fig.~\ref{fig:EScontour}.}
\label{fig:CCcontour}
\end{center}
\end{figure}

In Fig \ref{fig:CCcontour} we have plotted isocontours for the CC
day-night asymmetry for $\theta_{13}$ equal to 0, $9.2^\circ$ and
$12^\circ$ in order to observe the effect of a non-zero $\theta_{13}$
for the day-night asymmetry isocontours in the region of the LMA
solution.
Just as in the case of ES, the isocontours do not change dramatically
and the change is small compared to the uncertainty in the day-night
asymmetry. It is also apparent that the day-night asymmetry is smaller
for the ES detection than for the CC detection. This is to be
expected, since the ES detection is sensitive to the fluxes of
$\nu_\mu$ and $\nu_\tau$ as well as to the $\nu_e$ flux, while the CC
detection is only sensitive to the $\nu_e$ flux. The day-night
asymmetry for the best-fit values of Ref.~\cite{Maltoni:2003da} is
$A_{n-d} \simeq 4.7 \%$, which corresponds rather well to the values
presented in Refs.~\cite{deHolanda:2003nj,Bandyopadhyay:2003pk}. The
latest experimental value for the day-night asymmetry at SNO is
$A_{n-d} = (7.0 \pm 4.9 {}^{+1.3}_{-1.2})$ \% \cite{Ahmad:2002ka}.

\subsection{Determining $\boldsymbol{\theta_{13}}$ by using the 
day-night asymmetry}

As we observed earlier, the day-night asymmetry can, in principle, be
used for determining the mixing angle $\theta_{13}$ if the
experimental uncertainties in $\Delta m^2$, $\theta_{12}$, and the
day-night asymmetry itself were known to a larger accuracy. We now ask
how small the above uncertainties must be to obtain a reasonably low
uncertainty in $\theta_{13}$. To estimate the uncertainty $\delta
\theta_{13}$ in $\theta_{13}$, we use the pessimistic expression
\begin{eqnarray}
\delta\theta_{13} &\simeq& {\textstyle \left| 
\frac{\partial \theta_{13}}{\partial A_{n-d}}\right| \delta A_{n-d}
+ \left| \frac{\partial \theta_{13}}{\partial \theta_{12}}\right|
\delta \theta_{12}} \nonumber \\ &&+ {\textstyle \left| \frac{\partial
\theta_{13}}{\partial \Delta m^2}\right| \delta \Delta
m^2}. \label{eq:unc}
\end{eqnarray}

Let us suppose that, some time in the future, we have determined the
best-fit values for the parameters $\Delta m^2$ and $\theta_{12}$ to
be $\Delta m^2 = 6.9\cdot 10^{-5}$ eV$^2$ and $\theta_{12} =
33.2^\circ$ with the uncertainties $\delta\Delta m^2 \simeq 2.5\cdot
10^{-6}$ eV$^2$ and $\delta \theta_{12} \simeq
0.5^\circ$. Furthermore, suppose that we have measured the day-night
asymmetry for the ES reaction to be $A_{n-d} \simeq 0.03$ with an
uncertainty $\delta A_{n-d} \simeq 0.002$. These uncertainties roughly
correspond to one-tenth of the uncertainties of today's
measurements. Next, we estimate the partial derivatives of
Eq.~(\ref{eq:unc}) numerically and obtain ${\partial
\theta_{13}}/{\partial A_{n-d}} \simeq -{3000}^\circ$, ${\partial
\theta_{13}}/{\partial \theta_{12}} \simeq 0.4$, and ${\partial
\theta_{13}}/{\partial \Delta m^2} \simeq -2\cdot 10^{6}
{{}^\circ}/{{\rm eV}^2}$.  This gives an estimated error in
$\theta_{13}$ of $\delta \theta_{13} \simeq 12^\circ$ with $6.6^\circ$
from the uncertainty in $A_{n-d}$, $5.2^\circ$ from the uncertainty in
$\Delta m^2$ and $0.2^\circ$ from the uncertainty in
$\theta_{12}$. (Using the uncertainties $\delta\Delta m^2 \simeq
0.8\cdot 10^{-6}$ eV$^2$ and $\delta\theta_{12} \simeq 4^\circ$
suggested in Ref.~\cite{Bandyopadhyay:2003du}, we obtain the
uncertainty $\delta \theta_{13} \simeq 10^\circ$.) Thus, the
uncertainty in $\theta_{13}$ is less dependent on the uncertainty in
$\theta_{12}$ than the uncertainties in $A_{n-d}$ and $\Delta m^2$. We
observe that the precise value of the partial derivatives depend on
the point of evaluation, but that the values presented give a notion
about their magnitudes. If we use a more optimistic estimate instead
of Eq.~(\ref{eq:unc}), then $\delta \theta_{13}$ reduces to about
$8^\circ$. Unfortunately, this is still a quite large uncertainty and
in order to reach an uncertainty of about $1^\circ$, we need
measurements of $A_{n-d}$ and $\Delta m^2$ with uncertainties that are
about 100 times smaller than the uncertainties of today and a
measurement of $\theta_{12}$ with an uncertainty that is about 10
times smaller than today.

We would like to stress that the above can only be considered as an
indication of the uncertainties needed to determine $\theta_{13}$ and
serves mainly to present the idea of measuring $\theta_{13}$ with
the solar day-night asymmetry. In a realistic scenario, uncertainties
and fluctuations in the Earth matter density must be treated in a more
rigorous way than using a constant Earth potential; this would
probably require a full numerical simulation with three neutrino
flavors.

\section{Summary and conclusions}
\label{sec:S&C}

We have derived analytical expressions for the day $P_S$ and night
$P_{SE}$ survival probabilities for solar neutrinos in the case of
three neutrino flavors. The analytical result has been used to
numerically study the qualitative effect of a non-zero $\theta_{13}$
mixing on the day-night asymmetry $A_{n-d}$ at detectors. The
regenerative term in the three flavor framework was found to be
\begin{eqnarray}
P_{n-d} &=& -c_{13}^6 D_{3\nu} \frac{K V_E}{4a^2}
\sin^2(2\theta_{12})\sin^2(aL) \nonumber \\
&\simeq& -c_{13}^6 D_{3\nu} \frac{2E V_E}{\Delta m^2} 
\sin^2(2\theta_{12}) \nonumber \\
&&\times \sin^2\left({\textstyle \frac{\Delta m^2}{4E}L}\right), 
\end{eqnarray}
where $K = \Delta m^2 /(2E)$ and the approximation is good for
all parts of the LMA region. We have also noted that we regain the two
flavor expression for the regenerative term when $\theta_{13}
\rightarrow 0$. An important observation is that the quantitative
result depends linearly on the effective Earth electron density number
used to calculate the regenerative term. We have argued that the value
to be used is that of the Earth's crust.

In the study of the day-night asymmetry at detectors, it is apparent
that the relative effect of a non-zero $\theta_{13}$ in the LMA region
is increasing for increasing $\Delta m^2$ and decreasing $\theta_{12}$
and measured electron energy $T$. The dependence on $\theta_{12}$ is
also larger for smaller $T$ and larger $\Delta m^2$. This result holds
for both the elastic scattering and charged-current detection of
neutrinos.

We have also shown that the effects of a non-zero $\theta_{13}$ on the
isocontours of constant day-night asymmetry $A_{n-d}$ at detectors are
small compared to the current experimental uncertainty in
$A_{n-d}$. However, should this uncertainty and the uncertainties in
the fundamental parameters $\theta_{12}$ and $\Delta m^2$ become
much smaller in the future, the day-night asymmetry could be used to
determine $\theta_{13}$ as an alternative to future long baseline and
reactor experiments.

\section*{Acknowledgments}

We would like to thank Walter Winter for useful discussions and Thomas
Schwetz for providing the data of the allowed $\theta_{12}$-$\Delta
m^2$ parameter space.
This work was supported by the Swedish Research Council
(Vetenskapsr{\aa}det), Contract Nos.~621-2001-1611, 621-2002-3577
(T.O.), 621-2001-1978 (H.S.) and the G{\"o}ran Gustafsson Foundation
(G{\"o}ran Gustafssons Stiftelse) (T.O.).

\appendix

\section{The case of two neutrino flavors}
\label{ch:app}

In the two flavor case, we use the parametrization
\begin{equation}
U = \mtrx{cc}{c&s \\ -s&c},
\end{equation}
where $s = \sin \theta$ and $c = \cos \theta$ for the leptonic mixing
matrix. In this case, we can also write $P_{SE}$ as a function of
$P_S$ and $P_{2e}$ only such that
\begin{equation}
P_{n-d} = P_{SE}-P_S = \frac{1-2P_S}{\cos(2\theta)}(P_{2e}-\sin^2\theta).
\end{equation}

For $k_1$ and $k_2$, a general parametrization is
\begin{equation}
k_1 = \frac{1+D_{2\nu}}{2}, \ \ k_2 = \frac{1-D_{2\nu}}{2}.
\end{equation}
If we assume that all transitions among the matter eigenstates are
incoherent and/or negligible in magnitude, then $k_1$ and $k_2$ are
also given by
\begin{equation}
k_i = \int_0^{R_\odot} \d r f(r) 
[\cos^2 \hat\theta(r) P^s_{1i} + \sin^2 \hat\theta(r) P^s_{2i}],
\end{equation}
where $i \in \{1,2\}$, $P_{ki}^s$ is the probability of a neutrino in
the mass eigenstate $\ket{\nu_{M,k}}$ at position $r$ to exit the Sun
in the state $\ket{\nu_i}$, $\hat\theta$ is the mixing angle in
matter, and $f(r)$ is the normalized spatial distribution function of
the production of neutrinos. Using this, we obtain $D_{2\nu}$ as
\begin{equation}
D_{2\nu} = \int_0^{R_\odot} \d r f(r) \cos[2\hat\theta(r)] (1-2P_{\rm jump}),
\end{equation}
where $P_{\rm jump} = P_{12}^s = P_{21}^s$. From this expression, it
is easy to find that $|D_{2\nu}| \leq 1$, and thus, $0 \leq k_i \leq
1$, which is obviously a necessary condition. For the LMA solution of
the solar neutrino problem, we obtain $P_{\rm jump} < 10^{-1700}$ by
using a linear approximation of the effective potential at the point
of resonance. This is clearly negligible. The value of
$\cos(2\hat\theta)$ is calculated as
\begin{equation}
\label{eq:matangle}
\cos(2\hat\theta) = \frac{K\cos(2\theta) - 
V_{CC}}{\sqrt{[K\cos(2\theta) - V_{CC}]^2 + K^2 \sin^2(2\theta)}},
\end{equation}
where $V_{CC}$ is the effective matter density potential.
Now, the survival probability $P_S$ takes the simple form
\begin{equation}
P_S = \frac{1+D_{2\nu}\cos(2\theta)}{2}.
\end{equation}

For the propagation inside the Earth, we make the approximation that
the neutrinos traverse a sphere of constant electron number
density. This approximation is motivated by the fact that for current
detectors, most neutrinos do not pass through the Earth's core. As
long as neutrinos do not pass through the core, the only major
non-adiabatic point of the neutrino evolution is the entry into the
Earth's crust. As we will see, the regenerative term $P_{n-d} =
P_{SE}-P_S$ will oscillate quickly with $L$ and give an effective
averaging. It follows that the exact nature of the adiabatic process
inside the Earth does not matter as it only affects the frequency of
the oscillations. Thus, we may use any adiabatic electron number
density profile as long as we keep the electron number density at
entry into the Earth's crust and at detection fixed to the correct
values. In this case, the electron number density at entry into the
crust and at detection are the same, namely the electron number
density in the crust.

Keeping the electron number density constant, also the effective
electron neutrino potential is kept constant at $V_E = \sqrt 2 G_F
N_e$, where $G_F$ is the Fermi coupling constant and $N_e$ is the
electron number density. Exponentiating the Hamiltonian, we obtain the
time evolution operator and may calculate the probability
$P_{2e}$. The result of this calculation is
\begin{equation}
P_{2e} = \sin^2\theta + \frac{K V_E}{4a^2}\sin^2(2\theta)\sin^2(aL),
\end{equation}
where
\begin{equation}
a = \frac 12 \sqrt{K^2 - 2K V_E \cos(2\theta) + V_E^2}.
\end{equation}

Using the above results, we calculate $P_{n-d}$ and obtain the result
\begin{equation}
\label{eq:2Pnd}
P_{n-d} = -D_{2\nu}\frac{K V_E}{4a^2}\sin^2(2\theta)\sin^2(aL).
\end{equation}
For the allowed parameter space, the oscillation length in matter is
given by $L_{\rm osc} = 1/(2a) \simeq 1/K \sim 5\cdot 10^4$ m, which
is much shorter than the diameter of the Earth. Thus, we will have an
effective averaging of the term $\sin^2(aL)$ in Eq.~(\ref{eq:2Pnd}).

The expression for $P_{n-d}$ contains many expected features. For
example, when $V_E = 0$ then $P_{n-d} = 0$, the oscillation frequency
is just the one that we expect, and $P_{n-d} = 0$ for $D_{2\nu} = 0$
reflecting the fact that this corresponds to $k_1 = k_2 = 1/2$. In
Eq.~(\ref{eq:2Pnd}), we observe that the sign of $P_{n-d}$ depends on
the sign of $D_{2\nu}$, which depends on the point of production. If
we suppose that $P_{\rm jump} \ll 1$, then from
Eq.~(\ref{eq:matangle}) we find that $D_{2\nu}$ is negative if
$V_{CC}$ is larger than the resonance potential and positive if
$V_{CC}$ is smaller than the resonance potential. If the production
occurs near the resonance, then $D_{2\nu}$ will be small, since
$\hat\theta \simeq 45^\circ$ in this case. For the allowed parameter
space, the region of production for $^8$B neutrinos is such that
$V_{CC}$ is larger than the resonance potential, and thus, $D_{2\nu}$
will be negative. The same is true for $D_{3\nu}$.

\end{document}